\documentclass[runningheads]{llncs}
\usepackage[T1]{fontenc}
\usepackage{layouts}

\usepackage{graphicx}
\usepackage{amsmath, amssymb, amsfonts}
\usepackage{tikz}
\usetikzlibrary{calc}
\usepackage{cite}
\usepackage[section]{placeins}
\usepackage{color}
\usepackage{caption}
\usepackage{subcaption}
\usetikzlibrary{fit, positioning, arrows.meta}
\tikzset{neuron/.style={shape=circle, minimum size=1.5cm, 
		inner sep=0.2, draw, font=\small}, io/.style={shape=circle, minimum size=1.5cm,
		inner sep=0.2, draw, font=\small, fill=cyan!5}}
\usetikzlibrary{positioning, fit, arrows.meta, shapes}

\usepackage[utf8]{inputenc}
\usepackage[T1]{fontenc}
\usepackage{mathptmx}
\usepackage{etoolbox, comment,hyperref, amsmath}
\usepackage{soul}
\usepackage{lineno}
\usepackage{algorithm2e}
\usepackage{amsfonts}
\usepackage{bm}% bold math

\def\bx{{\boldsymbol{x}}}

\def\bu{{\boldsymbol{u}}}
\def\bU{{\boldsymbol{U}}}

\def\by{{\mathbf{y}}}
\def\rt{{\rm{t}}}

\def\bJ{{\boldsymbol{J}}}

\def\bQ{{\boldsymbol{Q}}}
\def\bR{{\boldsymbol{R}}}
\def\br{{\boldsymbol{r}}}

\usepackage{xcolor}

\begin{document}
	\title{Data-driven stability analysis of a chaotic time-delayed system  \thanks{This research has received financial support from the ERC Starting Grant No. PhyCo 949388.}}
	%
	%\titlerunning{Abbreviated paper title}
	% If the paper title is too long for the running head, you can set
	% an abbreviated paper title here
	%
	%\author{Georgios Margazoglou \inst{1} \orcidID{0000-0002-1374-9374} \and Luca Magri\inst{1, 2} \orcidID{0000-0002-0657-2611}}
	\author{Georgios Margazoglou \inst{1} \and Luca Magri\inst{1, 2, 3}}
	\authorrunning{G. Margazoglou and L. Magri}
	% First names are abbreviated in the running head.
	% If there are more than two authors, 'et al.' is used.
	%
	\institute{Department of Aeronautics, Imperial College London, London SW7 2AZ, UK  \and
		The Alan Turing Institute, London NW1 2DB, UK
		\and 
		Isaac Newton Institute for Mathematical Sciences, Cambridge, CB3 0EH, UK (visiting)
		\\ \email{g.margazoglou@imperial.ac.uk\, l.magri@imperial.ac.uk}}
	\maketitle              % typeset the header of the contribution
	\begin{abstract}		
		Systems with time-delayed chaotic dynamics are common in nature, from control theory to aeronautical propulsion. The overarching objective of this paper is to compute the stability properties of a chaotic dynamical system, which is time-delayed.  The stability analysis is based only on data. We employ the echo state network (ESN), a type of recurrent neural network, and train it on timeseries of a prototypical time-delayed nonlinear thermoacoustic system. By running the trained ESN  autonomously, we show that it can reproduce (i) the long-term statistics of the thermoacoustic system's variables, (ii) the physical portion of the Lyapunov spectrum, and (iii) the statistics of the finite-time Lyapunov exponents. This work opens up the possibility to infer stability properties of time-delayed systems from experimental observations.

	\end{abstract}

	\section{Introduction}\label{sec:intro}
	
	%The design of  aeroengines, such as gas-turbine or jet/rocket engines, requires a careful analysis to avoid thermoacoustic instabilities, among other possible malfunctions, which can lead to structural failure \cite{Magri2019}. In practice, thermoacoustic instabilities can develop when the thermal energy of the flame, that is converted into acoustic energy, exceeds dissipation mechanisms. This is typically encountered when the heat released by the flame is sufficiently in phase with the acoustic pressure\cite{Huhn_Magri_2020}. Hence, manufacturers design engines where small acoustic perturbations can decay quickly. To achieve this, a stability analysis is required in industrial preliminary design through parametric studies (most commonly eigenvalue analysis). That is because thermoacoustic systems can exhibit rich behaviours, such as fixed points, or periodic, quasi periodic and chaotic oscillations. Such behaviours are driven by bifurcations that occur through different design parameters and are connected with the geometrical features of the engine and the chemical composition of the flame. Hence, the optimization of engine design is a challenging task\cite{Magri2019}. In that regard, data-driven approaches have the potential to significantly improve and accelerate the preliminary engine design.
	
	Chaotic systems with time-delayed dynamics appear in a range of scientific fields	\cite{Lakshmanan2011dynamics}. Because of their dependence on both the present and past states, these systems have rich and intricate dynamics. Their chaotic behaviour can be assessed with stability analysis, which is a  mathematical tool that quantifies the system's response to infinitesimal perturbations. Stability analysis relies on the linearization of the time-delayed dynamical equations, which spawns the Jacobian of the system. From the Jacobian, we compute the Lyapunov Exponents (LEs), which are the key quantities to  quantifying  chaos \cite{Eckmann_Ruelle1985}.

	%	Modelling any physical system relies on assumptions and approximations. A different approach is to build model-free surrogate models from data, which has potential, given the abundance of experimental data. 
	%Data-driven approaches have proven powerful in various domains of society, science and technology. Still, for applications related to fluid dynamics, and depending on the context (in particular flow prediction), further improvements are required\cite{Brunton2020}. For flow prediction, that is partly because many neural network architectures struggle to learn the inherent chaotic nature of fluid dynamics. 
	A data-driven method, which considers the sequential nature of the dataset (e.g.~timeseries) to infer chaotic dynamics, is the recurrent neural network (RNN). Such networks have been successfully applied to learn chaotic dynamics (with no time delay) for different applications  \cite{Pathakchaos2017,Vlachas2020,Margazoglou2023}. The majority of RNNs require backpropagation through time for training, which can lead to vanishing or exploding gradients, as well as long training times \cite{Vlachas2020}. Instead, the echo state network (ESN), which we employ here, is trained via ridge regression, which eliminates backpropagation and provides a faster training \cite{Racca2021,Jaeger2004,Lukosevicius2012}. The objective of this paper is to train an ESN with data from a prototypical chaotic thermoacoustic system, which is a nonlinear time-delayed wave equation. 
	%Such systems describe the principal physical mechanisms of engines in aerospace engineering, such as gas-turbine or jet/rocket engines. Engine manufacturing requires a careful preliminary analysis to avoid thermoacoustic instabilities, among other possible malfunctions, which can lead to structural failure \cite{Magri2019}. In practice, thermoacoustic instabilities can develop when the thermal energy of the flame, that is converted into acoustic energy, exceeds dissipation mechanisms.
	We further assess the capabilities of the ESN to accurately learn the ergodic and stability properties of the thermoacoustic system, by calculating fundamental quantities, such as the Lyapunov exponents.
	We briefly review  stability analysis  for time-delayed systems in Sec.~\ref{sec:stability} and present the considered thermoacoustic system in Sec.~\ref{sec:rijke}. In Sec.~\ref{sec:ESN}, we discuss the ESN architecture and properties. We present the results in Sec.~\ref{sec:results} and conclude in Sec.~\ref{sec:Conclusion}.

	\section{Stability Analysis for time-delayed systems}\label{sec:stability}

	We consider a physical state $\bx(t) \in \mathbb{R}^D$, which is the solution of a nonlinear time-delayed dynamical system
	\begin{equation}
		\frac{d\bx}{dt} = f(\bx(t), \, \bx(t-\tau)),	\quad \bx(t)=\bx_0, \,\,\, \forall \,\,\, t\le0,
	\end{equation} 
	where $\tau$ is a constant time-delay. We analyse the system's stability by perturbing the state with infinitesimal  perturbations $\bu\sim\mathcal{O}(\epsilon), \;\epsilon\to 0$, as $\bx+\bu$, with $\bx\sim\mathcal{O}(1)$. Hence, we obtain the tangent linear equation
	\begin{equation}
		\frac{d\bU}{dt} = \bJ\left( \bx(t), \, \bx(t-\tau) \right)\bU,
	\end{equation}
	which involves the time-marching of $K\leq D$ tangent vectors, $\bu_i\in \mathbb{R}^D$, as columns of the matrix $\bU\in \mathbb{R}^{D\times K}$, $\bU = [\bu_1, \bu_2,\dots,\bu_K]$. This is a linear basis of the tangent space. The linear operator $\bJ  \in \mathbb{R}^{D\times D}$ is the Jacobian of the system, which is time-dependent in chaotic attractors. As shown in \cite{Margazoglou2023}, we can  extract the Jacobian of a reservoir computer by linearizing Eqs.~\eqref{eq:esn}. 
	
	We periodically orthonormalize the tangent space basis during time evolution by using a QR-decomposition of $\bU$, as  $\bU(t)=\bQ(t) \bR(t,\Delta t)$  and by updating the columns of $\bU$ with the columns of $\bQ$, i.e.~$\bU \leftarrow \bQ$ \cite{Eckmann_Ruelle1985}.  The matrix $\bR(t,\Delta t)\in \mathbb{R}^{K\times K}$ is upper-triangular and its diagonal elements $[\bR]_{i,i}$ are the local growth rates over a time span $\Delta t$ of $\bU$. The LEs are the time averages of the logarithms of the diagonal of $[\bR]_{i,i}$, i.e., 
	\begin{equation}
		\lambda_i = \lim\limits_{T\to\infty}\frac{1}{T} \int_{t_0}^{T}\ln [\bR(t,\Delta t)]_{i,i}dt.
		\label{eq:LEs}
	\end{equation}
	The FTLEs are defined as $\Lambda_i = \frac{1}{\Delta t} \ln [\bR]_{i,i}$, which quantify the expansion and contraction rates of the tangent space on ﬁnite-time intervals,  $\Delta t = t_2 - t_1$.

	\subsection{Time-delayed thermoacoustic system}\label{sec:rijke}
	As a practical application of time-delayed systems, we consider a thermoacoustic system, which is composed of three interacting subsystems, the acoustics, the flame and the hydrodynamics~(see, e.g., \cite{Magri2019}). 
	%The geometry of the acoustic resonator  and the chemical properties of the flame affect the system's stability properties. 
	The interaction of these sub systems can result in a positive feedback loop, which manifests itself as a thermoacoustic instability. If uncontrolled, this instability can lead to structural failure. %Hence, gas-turbine and rocket-motor manufacturers strive to optimize engine design to avoid such scenarios.
	We consider a prototypical time-delayed thermoacoustic system with a longitudinal acoustic cavity and a heat source modelled with a time-delayed model, following the same setup as in \cite{Magri2013,Huhn_Magri_2020,Huhn_Magri_2022}.
	The system is governed by the conservation of momentum, mass, and energy. Upon re-arrangement~\cite{Magri2019}, thermoacoustic dynamics are governed by the nondimensional  partial differential equations 
	%	\begin{equation}
		%		\begin{split}
			%			\frac{\partial u}{\partial t} &+	\frac{\partial p}{\partial x} = 0,\\
			%			\frac{\partial p}{\partial t} &+	\frac{\partial u}{\partial x} + \zeta p - \dot{q} \delta(x-x_f) = 0,
			%		\end{split}
		%		\label{eq:rijke_real}
		%	\end{equation}
	\begin{equation}
		\frac{\partial u}{\partial t} +	\frac{\partial p}{\partial x} = 0,\qquad
		\frac{\partial p}{\partial t} +	\frac{\partial u}{\partial x} + \zeta p - \dot{q} \delta(x-x_f) = 0,
		\label{eq:rijke_real}
	\end{equation}
	where $u$, $p$, $\dot{q}$, $x\in[0,1]$ and $t$ are the non-dimensional velocity, pressure, heat-release rate, axial coordinate and time, respectively; and   $\zeta$ is the damping coefficient, which takes into account all the acoustic dissipation. The heat source is assumed to be small compared to the acoustic wavelength, and it is modelled as a point in the grid, via the Dirac delta distribution $\delta(x-x_f)$, located at $x_f=0.2$. The heat-release rate is provided by a modified King's law, $\dot{q}(t) = \beta \left(\sqrt{|1+u_f(t-\tau)|} - 1 \right)$, which is a nonlinear time-delayed model. For the numerical studies of this paper, we set $\beta=7.0$ for the heat parameter, and $\tau=0.2$ for the time delay. Those values ensure chaotic evolution, and encapsulate all information about the heat source, base velocity and ambient conditions. 
	%A high-accuracy stability analysis for the system's dynamics at changing $\beta$ and $\tau$ was presented in \cite{Huhn_Magri_2020} and a data-assimilation strategy for this system in \cite{Traverso2019}.
	%  Note that Eqs.~\eqref{eq:rijke_real} are derived by linearising the inviscid momentum and energy equations based on three assumptions. First, the acoustics are small perturbations onto a mean flow at rest with uniform density. Second, both viscosity and diffusivity are negligible. Third, the acoustics are one-dimensional, i.e. the cut-on frequency of the tube is much higher than the frequency of the instability. Also, the acoustic pressure at the boundaries is zero, as the tube has open ends.

	As in~\cite{Huhn_Magri_2020}, we transform the time-delayed problem into an initial value problem. This is mathematically achieved by modelling the advection of a perturbation $v$ with velocity $\tau^{-1}$ as
	\begin{equation}
		\frac{\partial v}{\partial t} + \frac{1}{\tau}\frac{\partial v}{\partial X} = 0,\quad 0 \le X\le 1, \quad	v(X=0,t) = u_f(t).
		\label{eq:advection}
	\end{equation}
	We discretise Eqs.~\eqref{eq:rijke_real} by a Galerkin method. First, we separate the acoustic variables in time and space as $u(x,t) = \sum_{j=1}^{N_g} \eta_j(t)\cos(j \pi x)$, and  $p(x,t) =  -\sum_{j=1}^{N_g} \mu_j(t) \sin(j \pi x)$, in which the spatial functions are the acoustic eigenfunctions of the configuration under investigation. %Hence, each spatial function is a natural acoustic mode of the open-ended tube.
	Then, we project Eqs.~\eqref{eq:rijke_real} onto the Galerkin spatial basis $\{ \cos(\pi x),$ $\cos(2 \pi x),$ $\dots, \cos(N_g \pi x)\}$ to obtain
	\begin{equation}
		\dot{\eta}_j - j \pi \mu_j = 0, \qquad
		\dot{\mu}_j + j \pi \eta_j + \zeta_j \mu_j + 2 \dot{q} \sin(j \pi x_f) = 0.
		\label{eq:rijke_galerkin}
	\end{equation}
	The system has $2N_g$ degrees of freedom. 
	The time-delayed velocity becomes $u_f(t-\tau) = \sum_{k=1}^{N_g} \eta_k(t-\tau) \cos(k \pi x_f)$,
	and the damping, $\zeta_j$, is modelled by % a modal expression that damps out higher-frequency oscillations,
	$\zeta_j = c_1 j^2 + c_2 j^{1/2}$, where $c_1=0.1$ and $c_2=0.06$. The equation for linear advection, Eq.~\eqref{eq:advection}, is discretised using $N_c+1$ points with a Chebyshev spectral method. This discretisation adds $N_c$ degrees of freedom, thus a total of $D=2N_g+N_c=30$ in our case, as $N_g=N_c=10$. We integrate Eqs.~\eqref{eq:rijke_galerkin} with a fourth order Runge-Kutta scheme and timestep $dt=0.01$.
	
	%\lm{[I would shorten this section on some physics details by, say, 30\% to 50\%. Cite also Traverso in ICCS and Magri 2013 JFM please.]}

	\section{Echo State Network}\label{sec:ESN}
	By applying the method of~\cite{Margazoglou2023} to time-delayed problems, we linearize the Echo State Network (ESN) \cite{Jaeger2004} to calculate the stability properties of  chaotic systems. ESNs are proven effective for accurate learning of chaotic dynamics (see e.g.~\cite{Pathakchaos2017,Vlachas2020,Huhn2020proc,Doan2021_prsa,Racca2021,Margazoglou2023}). The ESN is a reservoir computer \cite{Jaeger2004}. It has a sparsely-connected single-layer hidden state, which is termed  ``reservoir''. The reservoir weights, $\mathbf{W}$, as well as the input-to-reservoir weights, $\mathbf{W}_{\mathrm{in}}$, are randomly assigned and remain fixed through training and testing. The reservoir-to-output weights, $\mathbf{W}_{\mathrm{out}}$, are trained via ridge regression. The evolution equations of the reservoir and output are,  respectively  
	\begin{equation}
		\br(\rt_{i+1}) = \tanh\left([\hat{\by}_{\mathrm{in}}(\rt_{i});b_\mathrm{in}]^T\mathbf{W}_{\mathrm{in}}+\br(\rt_{i})^T\mathbf{W}\right), \qquad \by_{\mathrm{p}}(\rt_{i+1}) = [\br(\rt_{i+1});1]^T\mathbf{W}_{\mathrm{out}},
		\label{eq:esn}
	\end{equation}
	where at any discrete time $\rt_i$ the input vector, $\by_{\mathrm{in}}(\rt_{i}) \in \mathbb{R}^{N_y}$, is mapped into the reservoir state $\br \in \mathbb{R}^{N_r}$, by the input matrix, $\mathbf{W}_{\mathrm{in}}$, where $N_r \gg N_y$\cite{Jaeger2004,Racca2021}. Here, $\hat{(\;\;)}$ indicates normalization by the component-wise maximum-minus-minimum range of the target in training set, $^T$ indicates matrix transposition, and the semicolon indicates array concatenation.
	The dimensions of the weight matrices are $\mathbf{W}_{\mathrm{in}} \in \mathbb{R}^{(N_y+1)\times N_r }$, $\mathbf{W} \in \mathbb{R}^{N_r\times N_r}$ and $\mathbf{W}_{\mathrm{out}} \in \mathbb{R}^{(N_{r}+1)\times N_y}$. The hyperparameter input bias, $b_{\mathrm{in}}=1$, is  selected to have the same order of magnitude as the normalized inputs, $\mathbf{\hat{y}}_{\mathrm{in}}$. The dimensions of the input and output vectors are equal to the dimension of the  dynamical system; here described by Eqs.~\eqref{eq:rijke_galerkin}, i.e.~$N_y \equiv D$. 
	Furthermore, $\mathbf{W}_{\mathrm{out}}$ is trained via the minimization of the mean square error $\textrm{MSE} = \frac{1}{N_{\mathrm{tr}}N_y} \sum_{i=0}^{N_{\mathrm{tr}}} || \by_{\mathrm{p}}(\rt_{i}) - \by_{\mathrm{in}}(\rt_{i})||^2$ between the outputs and the data over the training set, where $||\cdot||$ is the $L_2$ norm, $N_{\mathrm{tr}}+1$ is the total number of data in the training set, and $\by_{\mathrm{in}}$ the input data on which the ESN is trained.
	
	Training the ESN is performed by solving with respect to $\mathbf{W}_{\mathrm{out}}$ via ridge regression of the equation $(\mathbf{R}\mathbf{R}^T + \beta \mathbb{I})\mathbf{W}_{\mathrm{out}} = \mathbf{R} \mathbf{Y}_{\mathrm{d}}^T$. In the previous expression $\mathbf{R}\in\mathbb{R}^{(N_r+1)\times N_{\mathrm{tr}}}$ and $\mathbf{Y}_{\mathrm{d}}\in\mathbb{R}^{N_y\times N_{\mathrm{tr}}}$ are the horizontal concatenation of the reservoir states with bias, $[\br(\rt_{i});1]$, $\rt_i \in [0,T_{\mathrm{train}}]$, and of the output data, respectively; $\mathbb{I}$ is the identity matrix and $\beta$ is the Tikhonov regularization parameter \cite{Tikhonov1995}. Therefore the ESN does not require backpropagation. The ESN can run in two configurations, either open-loop or closed-loop. In open-loop, which is necessary for the training stage, the input data is given at each step, allowing for the calculation of the reservoir timeseries $\br(\rt_{i})$, $\rt_{i} \in [0,T_{\mathrm{train}}]$. In closed-loop the output $\by_{\mathrm{p}}$ at time step $\rt_{i}$, is recurrently used as an input at time step $\rt_{i+1}$, allowing for the autonomous temporal evolution of the network. The closed-loop configuration is used for validation (i.e. hyperparameter tuning) and testing, but not for training. 
	
	Regarding validation, we use the chaotic recycle validation (RVC), as introduced in \cite{Racca2021}. It has proven to be a robust strategy, providing enhanced performance of the ESN, compared to standard strategies, as recently successfully applied in \cite{Racca2022,Margazoglou2023}. Briefly, in RVC the network is trained only once on the entire training dataset (in open-loop), and validation is performed on multiple intervals already used for training (but now in closed-loop). The validation interval simply shifts as a small multiple of the first Lyapunov exponent, $N_{\mathrm{val}}=3\lambda_{1}$ here. 
	The key hyperparameters that we tune are the input scaling $\sigma_{\mathrm{in}}$ of the input matrix $\mathbf{W}_{\mathrm{in}}$, the spectral radius $\rho$ of the matrix $\mathbf{W}$, and the Tikhonov parameter $\beta$. %The input scaling defines the range $[-\sigma_{\mathrm{in}},\sigma_{\mathrm{in}}]$ that we sample from a uniform distribution to fill a randomly chosen element per row of the input matrix $\mathbf{W}_{\mathrm{in}}$, with the rest being set to zero. 
	Furthermore, $\sigma_{\mathrm{in}}$ and $\rho$ are tuned via Bayesian Optimization in the hyperparameter space $[\sigma_{\mathrm{in}},\rho]=[0.1,5]\times[0.1,1]$ in logarithmic scale, while for $\beta$ we perform a grid search $\{10^{-6},10^{-8}, 10^{-10},10^{-12}\}$ within the optimal  $[\sigma_{\mathrm{in}},\rho]$. The reservoir size is $N_r=400$.
	The connectivity of matrix $\mathbf{W}$ is set to $d=3$. We further add to the training and validation data a Gaussian noise with zero mean and standard deviation, $\sigma_n=0.0006\sigma_y$, where $\sigma_y$ is the standard deviation of the data component-wise (noise regularizes the problem, see  \cite{Vlachas2020,Racca2021, Margazoglou2023} for more details). The ESN is trained on a training set (open-loop) of size $200\tau_\lambda$, and is tested on a test set (closed-loop) of size $4000\tau_\lambda$, where $\tau_\lambda=1/\lambda_1$ is the Lyapunov time, which is the inverse of the maximal Lyapunov exponent $\lambda_1\approx0.13$ in our case.% Much larger values of $N_r>1000$ led to underestimation of $\lambda^{\rm ESN}_1$, although the statistics of the variables were always very good (see Fig.~\ref{fig:1}). Instead, smaller values, $N_r<200$, led to a harder learning of the thermoacoustic system's chaotic nature.

	\section{Results}\label{sec:results}
	We analyse the statistics produced by the autonomous temporal evolution of the ESN and the target time-delayed system. The selected observables are the statistics of the system's chaotic variables, the Lyapunov exponents (LEs), and the statistics of the finite-time Lyapunov exponents (FTLEs).

\begin{figure}[!h]
		\centering
		\includegraphics[width=.99\linewidth]{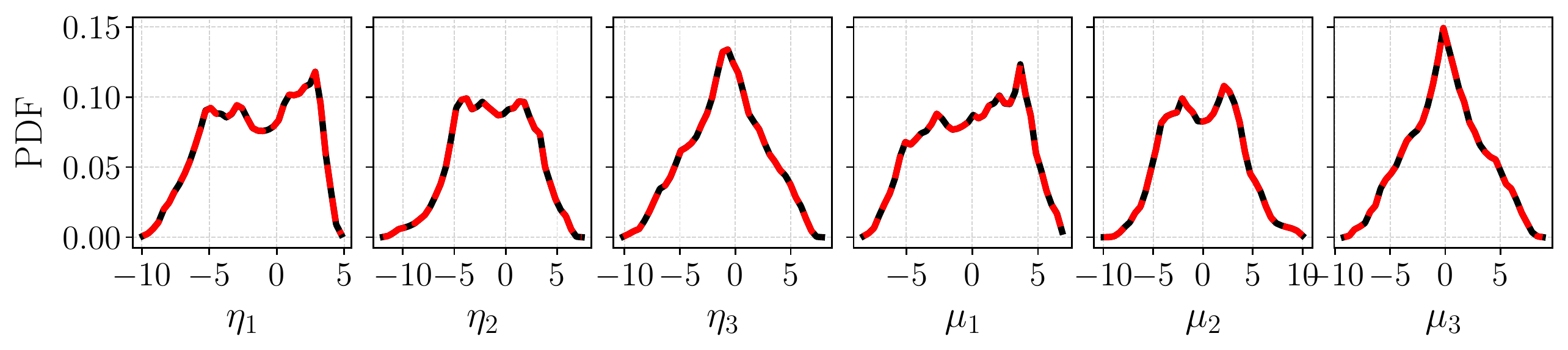}
		\caption{Probability density functions (via histograms) of the three first Galerkin modes, $\eta_i$, $\mu_i$, $i={1,2,3}$. Black is for target and dashed red for ESN. The statistics for ESN are collected in autonomous evolution on unseen data, after training and validation.}
		\label{fig:1} 
	\end{figure}

	First, we test the capabilities of the ESN to learn the long-term statistical properties of the thermoacoustic system by measuring the probability density function (PDF) of the learned variables, $\by$. In Fig.~\ref{fig:1}, we show the PDF of the first three components of the Galerkin modes (see Eq.~\eqref{eq:rijke_galerkin})  $\eta_i$, $\mu_i$ for $i={1,2,3}$, in which the black line corresponds to the target and the dashed red to the ESN. The ESN predictions are in agreement with the target, including the variables that are not shown here. 
	
\begin{figure}[!h]
		\centering
		\includegraphics[width=.6\linewidth]{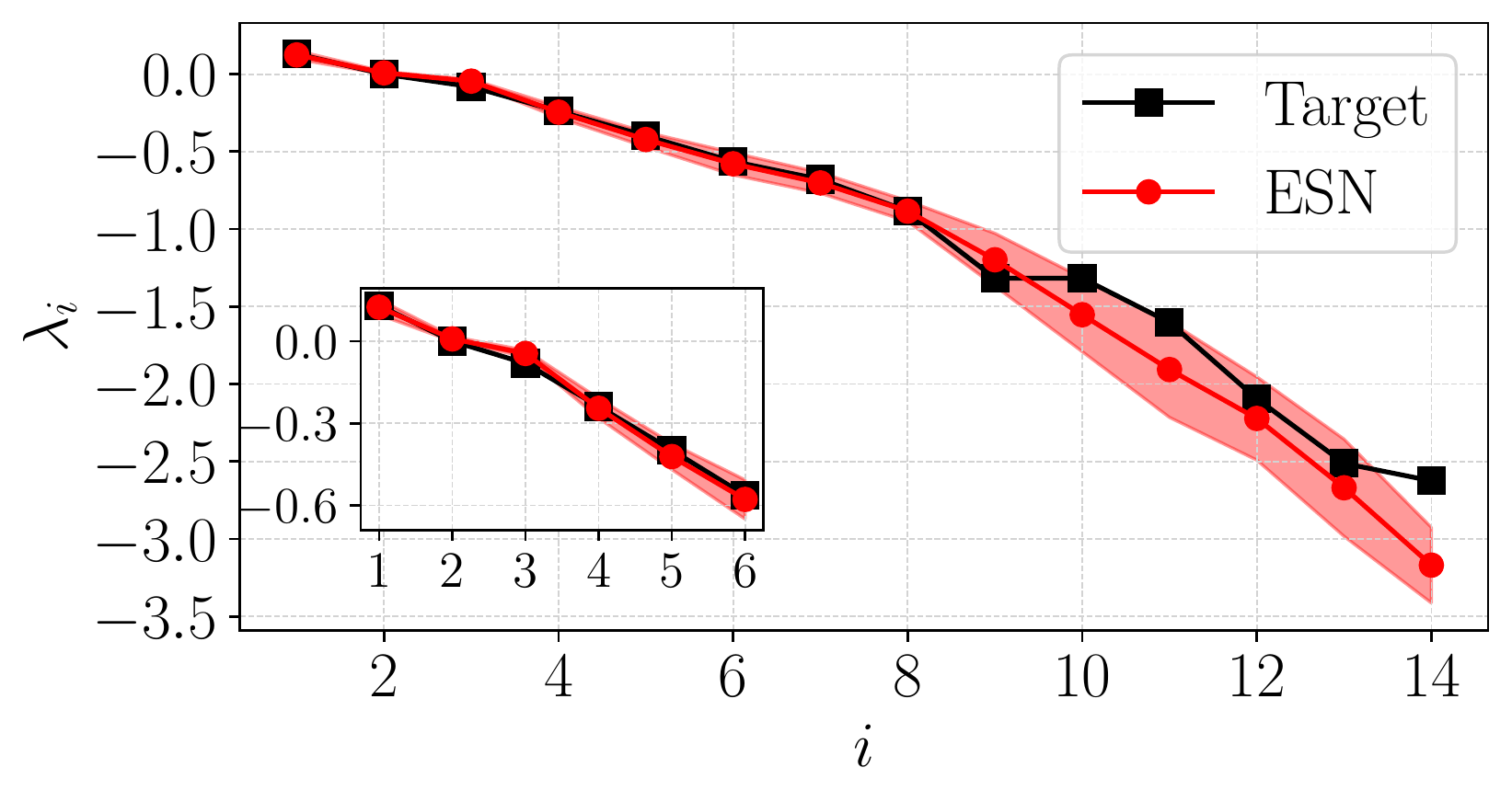}
		\caption{The first 14 (outer plot) and the first 6 (inner plot) Lyapunov exponents. The shaded red region indicates the error based on the ensemble of 5 ESNs. The statistics for ESN are collected in closed-loop mode.}
		\label{fig:2} 
	\end{figure}

	Testing the accuracy of the calculated LEs in autonomous evolution is a harder consistency check for the ESN. Indeed, the ESN has been trained only on timeseries of the variables $\eta_i$, $\mu_i$, and $v$. Therefore, a good agreement of the LEs means that the ESN is capable to accurately reproduce intrinsic chaotic properties of the system's attractor. The LEs of the ESN are calculated following \cite{Margazoglou2023}, and in Fig.~\ref{fig:2} we compare the first $K=14$ LEs. We also add an inner plot showing, in additional detail, the first 6 LEs. Each LE is the average of the measured LEs from five selected independent ESNs used for the analysis. We train the ESNs independently on different chaotic target sets. The shaded region corresponds to the standard deviation per $\lambda_i$ from those five ESNs. 
	There is close agreement for the first 8 exponents. In particular, we measure the leading, and only positive, exponent $\lambda^{\rm targ}_1\approx0.130$ and $\lambda^{\rm ESN}_1\approx0.124$ for ESN, which gives a 4.7\% absolute error. The ESN also provides an accurate estimate of the neutral exponent ($\lambda_2=0$) with $\lambda^{\rm ESN}_2\approx0.008$. The rest of the exponents, $\lambda_{i}, \,i\ge3$, are negative and the ESN achieves a small $8.2\%$ mean absolute percentage error for all. Note that a gradual disagreement of the negative exponents, which are sensitive due to the numerical method, between ESN and target has also been reported in \cite{Pathakchaos2017,Vlachas2020} for the one-dimensional Kuramoto-Sivashinsky equation.  
	
\begin{figure}[!h]
		\centering
		\includegraphics[width=.99\linewidth]{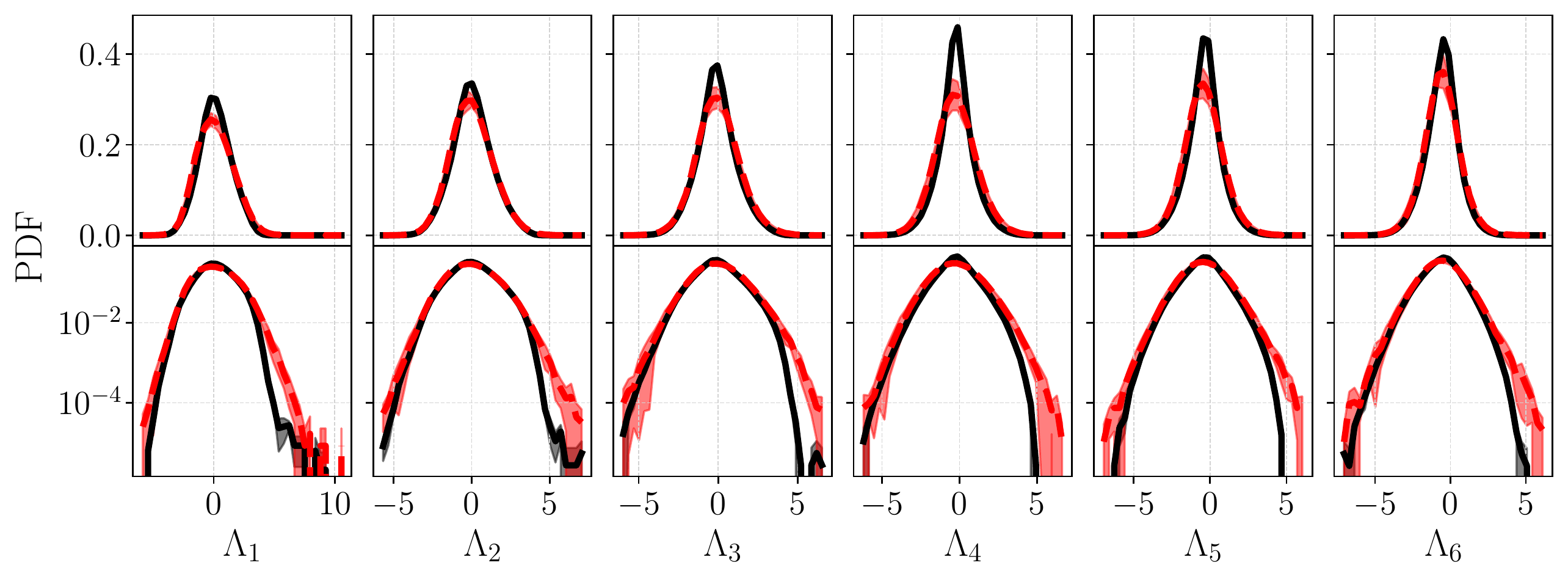}
		\caption{Probability density functions of the first six finite-time Lyapunov exponents. Black is for target and dashed red for ESN. The y-axis in bottom row is in logarithmic scale. }
		\label{fig:3} 
	\end{figure}

	Figure~\ref{fig:3} shows the PDF of the first six FTLEs of the target (black line) and ESN (dashed red line). We collect the statistics from 5 independent ESNs, thus creating 5 histograms of FTLEs.
	We then average those histograms bin-wise, and the standard deviation of each averaged bin is given by the shaded regions, which are more pronounced at the tails. In Fig.~\ref{fig:3} the plots in each column of bottom row are identical to the ones in upper row, with the difference that the y-axis is in logarithmic scale to emphasize the agreement of the statistics also at the tails, as the agreement for the most probable statistics close to the peak is  good (upper row). Note that the mean of each PDF should correspond to each Lyapunov exponent $\lambda_i$ (i.e.~of Fig.~\ref{fig:2}), which is indeed the case.
	
	We also report the values for the Kaplan-Yorke dimension for both ESN and target. This dimension is an upper bound of the attractor's fractal dimension\cite{Eckmann_Ruelle1985}. It is given by $D_{\text{KY}} = k + \frac{\sum_{i=1}^{k} \lambda_{i}}{|\lambda_{i+1}|}$, where $k$ is such that the sum of the first $k$ LEs is positive and the sum of the first $k+1$ LEs is negative. We obtain $D^{\rm targ}_{\text{KY}}\approx3.37$ for target and  $D^{\rm ESN}_{\text{KY}}\approx3.46$ for ESN, which results in a  2.7\% absolute error. This observation further confirms the ability of the ESN to accurately learn the properties of the chaotic attractor.
	
	\section{Conclusion}\label{sec:Conclusion}
	%We propose a method for computing the stability properties of chaotic solutions of time-delayed systems from data only. 
	%We employ the echo state network (ESN) as a surrogate model for  learning the chaotic dynamics from timeseries of observations of a time-delayed system, and computing the long-term statistical and stability properties.
	We propose a method to compute the stability properties of chaotic solutions in time-delayed systems using only data. We use the echo state network (ESN) as a surrogate model for learning the chaotic dynamics from the time series observations of the system and determining its long-term statistical and stability properties.
	 By considering the ESN as a discrete dynamical system, we linearize the map \eqref{eq:esn} to derive the tangent evolution of the attractor through the Jacobian. When running the  ESN in a long autonomous mode (closed-loop), 
	we show that (i) the long-term statistics of the variables are correctly learned, (ii) the physical portion of the Lyapunov spectrum is correctly predicted, and (iii) the finite-time Lyapunov exponents and their statistics are correctly inferred. This work opens up the possibility to infer the stability of nonlinear and time-delayed dynamical systems from data.

	\bibliographystyle{splncs04}
	\bibliography{bibliography}
\end{document}